\input phyzzx
\sequentialequations
\overfullrule=0pt
\tolerance=5000
\nopubblock
\twelvepoint

\line{\hfill }
\line{\hfill IASSNS 96/95}
\line{\hfill hep-th/9609099 }
\line{\hfill September 1996}

\titlepage
\title{Asymptotic Freedom\foot{Lecture on receipt of the Dirac medal
for 1994, October 1994.}}

\vskip .2cm
\author{Frank Wilczek\foot{Research supported in part by DOE grant
DE-FG02-90ER40542.~~~wilczek@sns.ias.edu}}
\vskip.2cm
\centerline{{\it School of Natural Sciences}}
\centerline{{\it Institute for Advanced Study}}
\centerline{{\it Olden Lane}}
\centerline{{\it Princeton, N.J. 08540}}
 
\endpage
 
\abstract{I discuss how the basic phenomenon
of asymptotic freedom in QCD can be understood in elementary
physical terms.  Similarly, I discuss how the long-predicted
phenomenon of ``gluonization of
the
proton'' -- recently spectacularly confirmed at HERA -- 
is a rather direct manifestation of the physics of
asymptotic freedom.
I review the broader significance of asymptotic freedom in QCD in
fundamental physics: how on the one hand it guides the interpretation
and now even the design of experiments, and how on the other it
makes possible a rational, quantitative theoretical approach to
problems of unification and early universe cosmology.}

\endpage

\REF\landau{L. Landau, in {\it Niels Bohr and the
Development of Physics}, ed. W. Pauli.  (McGraw-Hill,
New York 1955).}

\REF\weak{S. Weinberg, {\it Phys. Rev. Lett. } {\bf 19}, 1264 (1967);
A. Salam, in {\it Elementary Particle Physics}, ed. N. Svartholm
(Almqvist and Wiksells, Stockholm, 1968) p. 367; S. Glashow, J.
Iliopoulos, 
and L. Maiani, {\it Phys. Rev.} {\bf D2} 1285 (1971).}

\REF\tHren{G. 't Hooft, {\it Nucl. Phys.} {\bf B35}, 167 (1971).}

\REF\partons{R. Feynman, {\it Phys. Rev. Lett.\/}
{\bf 23} 1415 (1969);
J. Bjorken and E. Paschos {\it Phys. Rev.\/} {\bf 185} 1975 (1969).}

\REF\grosswil{D. Gross and F. Wilczek, {\it Phys. Rev. Lett}. {\bf
30}, 1343 (1973).}

\REF\pol{H. D. Politzer, {\it Phys. Rev. Lett}. {\bf 30}, 1346 (1973).}

\REF\nielsen{N. K. Nielsen, {\it Am. J. Phys.} {\bf 49}, 1171 (1981).}

\REF\hughes{R. Hughes, {\it Nucl. Phys.}  {\bf B186}, 376 (1981).}

\REF\zee{A. Zee, {\it Phys. Rev.}  {\bf D7}, 3630 (1973).}

\REF\colegross{S. Coleman and D. Gross, {\it Phys. Rev. Lett.}
{\bf 31}, 851 (1973).}

\REF\metals{See for example Ziman, {\it Principles of the Theory of
Metals}, (Cambridge University Press, 1972)
pp. 330-332.}

\REF\swein{S. Weinberg, in {\it Lectures on Elementary Particles and
Quantum Field Theory\/} S. Deser, M. Grisaru, and H. Pendleton, eds.
p. 324 (MIT Press, Cambridge, 1970).}

\REF\kramers{The general philosophy of this program may have been
first expressed by
Kramers.  Its famous implementation in quantum electrodynamics is
documented in the following reference, in which the contribution by 
F. Dyson, {\it Phys. Rev.} {\bf 75}, 486 (1949) is especially noteworthy.}

\REF\schwinger{{\it Selected Papers on Quantum Electrodynamics}, 
J. Schwinger, ed. (Dover, New York 1958).}

\REF\euheis{Indeed what we would now call the effective action for a
constant gauge field was derived by W. Heisenberg and H. Euler,
{\it Z. Phys.} {\bf 98}, 714 (1936), 
and
explicitly interpreted as polarizability of the vacuum; and the
concept of running coupling was introduced by M. Gell-Mann and F. Low, 
{\it Phys. Rev.} {\bf 95}, 1300 (1954);   
and in a slightly different form vigorously discussed by Landau and his
school in the 1950s.}

\REF\proptime{J. Schwinger, {\it Phys. Rev.} {\bf 82}, 664 (1951).}

\REF\fkt{See for example the historic presentations by W. Panofsky, 
{\it Proc. 14th Int. Conf. on High Energy Physics (Vienna)}, 
J. Prentki and J. Steinberger, ed. CERN, Geneva; and G. Miller 
{\it et al}. {\it Phys. Rev.} {\bf D6}, 3011 (1972).}

\REF\bjlim{J. Bjorken, {\it Phys. Rev.} {\bf 179}, 1547 (1969).}

\REF\gwtwo{D. Gross and F. Wilczek, {\it Phys. Rev.} {\bf D9}, 980 (1974).}

\REF\gp{H. Georgi and H. Politzer, {\it Phys. Rev.} {\bf D9}, 416
(1974).}

\REF\wope{K. Wilson, {\it Phys. Rev.} {\bf 179}, 1499 (1969); 
{\it Phys. Rev.} {\bf D3}, 1818 (1971).}

\REF\gwone{The results that follow from 
this sort of analysis were first described in
D. Gross and F. Wilczek, {\it Phys. Rev.} {\bf D8}, 3633 (1973).}

\REF\cg{C. Callan and D. Gross, {\it Phys. Rev. Lett.} {\bf 22}, 156 (1969).}

\REF\sixman{A. de Rujula, S. Glashow, H. Politzer, S. Treiman, F.
Wilczek,
and A. Zee, {\it Phys. Rev.} {\bf D10}, 1649 (1974).}

\REF\zwt{A. Zee, F. Wilczek, and S. Treiman, {\it Phys. Rev.} 
{\bf D10},  2881 (1974).}

\REF\bffig{R. Ball and S. Forte, hep-ph/9409373; see also the
following reference.}

\REF\bfcalc{A large body of work is
nicely summarized in R. Ball and S. Forte, hep-ph/9512208, to be published in 
proceedings of the XXXV Cracow School of Theoretical Physics.}

\REF\altpar{G. Altarelli and G. Parisi, {\it Nucl. Phys.}
{\bf B126}, 298 (1977).}

\REF\ks{J. Kogut and L. Susskind, {\it Physics Reports\/} {\bf 8C}, 76 (1973).}

\REF\datareviews{See for example {\it QCD--20 Years Later}, eds. P. Zerwas
and H. Kastrup, (World Scientific, Singapore, 1992); Particle Physics:
 Perspectives and Opportunities, eds. R. Peccei and others (World
 Scientific, Singapore 1995), and references therein.}

\REF\incljet{Data from the CDF Collaboration, 1989 (Web Site
 http://www-cdf.fnal.gov/). }

\REF\antenna{Data from the OPAL Collaboration, (Web
Sitehttp://www1.cern.ch/Opal/). }

\REF\heavy{Figure 4 taken from J. Shigemitsu, {\it Quarkonium Physics
 and $\alpha_{strong}$ from Quarkonia}, hep-lat/9608058.}

\REF\running{ Figure 5 taken from M. Schmelling, talk given at the XV International
Conference on Pysics in Collision, Cracow, Poland, June 1995,
CERN-PPE/95-29 (Aug 1995).}

\REF\creutz{An excellent review of the principles of
lattice gauge theory, by the theorist who first carried
through the argument just mentioned, is M. Creutz,
{\it Quarks, Gluons,
and Lattices\/} (Cambridge, 1983).}

\REF\largeN{G. 'tHooft, {\it Nucl. Phys.\/} {\bf B72},
461 (1974); E. Witten, {\it Nucl. Phys/\/} {\bf B160},
57 (1979).}

\REF\tHooft{G. 'tHooft, {\it Phys. Rev. Lett.\/}
{\bf 37}, 8 (1976); C. Callan, R. Dashen, and D. Gross {\it
Phys. Lett/\/} {\bf 63B}, 334 (1976); R. Jackiw and
C. Rebbi, {\it Phys. Rev. Lett.\/} {\bf 37}, 172 (1976).}

\REF\pq{R. Peccei and H. Quinn, {\it Phys. Rev. Lett.\/}
{\bf 38}, 1440 (1977).}

\REF\axion{S. Weinberg, {\it Phys. Rev. Lett.\/}
{\bf 40}, 223 (1978); F. Wilczek, {\it Phys. Rev. Lett.\/}
{\bf 40}, 279 (1978).}

\REF\gg{H. Georgi and S. Glashow, {\it Phys. Rev. Lett.\/}
{\bf 32}, 438 (1974).}

\REF\gqw{H. Georgi, H. Quinn, and S. Weinberg, {\it Phys. Rev.
Lett.\/} {\bf 33}, 451 (1974).}

\REF\dim{For a fuller discussion,
including extensive references, see
S. Dimopoulos, S. Raby, and F. Wilczek,
{\it Physics Today\/} {\bf 44}, October, p. 25  (1991)}

\REF\wein{S. Weinberg, {\it Gravitation and Cosmology\/}
(Wiley, New York 1972).}

I am very pleased to accept your award today.  On this occasion I
think it is appropriate to discuss with you the circle of ideas around
asymptotic
freedom.  
After a few remarks about its setting in intellectual history,
I will begin by explaining the physical origin of asymptotic
freedom in QCD; then I will show how a recent, spectacular
experimental observation -- the `gluonization' of the proton --
both confirms and illuminates its essential nature; then I will
discuss some of its broader implications for fundamental physics.

It may be difficult for
young people who missed experiencing it,
or older people with fading
memories, fully to imagine the intellectual atmosphere 
surrounding the strong interaction in the
1960s and early 1970s.  It is quite
instructive to look into the literature of those times.
Many if not most
theoretical papers dealing with the strong interaction
contained an obligatory ritual mantra wherein the S-matrix or
bootstrap was invoked, before getting down to their actual point
(often rather tenuously connected to those theological principles).
Use of strict
quantum field theory was considered to be naive, in rather poor taste,
an occasion for
apology.
One hears an echo of these attitudes
even in the conclusion of Gell-Mann's
famous 1972 summary talk at the NAL conference:

``Let us end by emphasizing our main point, that it may well be
possible to construct an explicit theory of hadrons, based on quarks
and some kind of glue, treated as fictitious, but with enough
physical properties abstracted and applied to real hadrons to
constitute a complete theory.  Since the entities we start with are
fictitious, there is no need for any conflict with the bootstrap or
conventional dual parton point of view.''

What were the reasons for this suspicion
of quantum field theory,
which in retrospect appears strange?  Part of the
reason was historical.  The late 1940s and early 50s saw
what appeared on the face of it to be a great triumph for
quantum field theory, the triumph
of renormalization theory in QED.  However
the procedures developed at that time
for solving, or even making sense of,
the
equations of QED
were intrinsically tied to a perturbative
expansion in powers of the coupling constant.  For
QED this coupling is indeed small, but in the
then-current candidate
quantum field theory of the strong interactions, Yukawa's
$\pi$ meson theory, it was clear that the coupling would have
to be
large for the theory
to have any chance of agreeing with experiment.  Thus although
this theory was not known to be wrong, it was certainly
useless in practice.  Attempts to solve the theory without
resorting to perturbation theory did not succeed, both for
practical reasons and for a fundamental one
that we will discuss momentarily.

As the rich phenomenology of resonance physics was discovered,
theorists for the most part
made progress toward digesting it not by the top-down
approach of deriving mathematical consequences from a powerful
fundamental theory, but rather by more modest methods based on
symmetry and high-class kinematics.  (In
the category of high-class kinematics I
include dispersion relations, derived from causality, and
S-matrix
model-building guided by pole-dominance or
narrow-resonance approximations together with the
constraint of unitarity.)

Thus quantum field theory gradually lost much of its luster.
The successes of field theory in QED were rationalized as
due to
a lucky accident.  One could to a certain extent recover
these successes from the less committal point of view
fashionable
in strong interaction physics, along the following lines:
the weak coupling expansion of
quantum field theory is essentially a systematic way of unitarizing
the single pole amplitude for photon exchange, supplemented with
the assumption that the relevant dispersion relations need no
subtraction.
This philosophy appeared especially sensible given that renormalization
theory failed even for the other available weak-coupling theories
of the weak interactions and of gravitation, while the modest
semi-kinematic approach worked perfectly well in these domains,
and was extremely fruitful
in untangling the weak interactions of hadrons.

But the difficulties in accepting quantum field
theory at face value were not only matters of history and
sociology.  The only powerful method for
extracting consequences
from non-trivial interacting quantum field theories
was perturbation theory in the coupling.  This perturbation
theory, implemented in a straightforward way, gave infinite
results order by order due to the exchange of highly virtual
quanta.  Tomonoga, Schwinger, and Feynman, building on
qualitative insights of Kramers and Bethe, were able to
make sense of the perturbation theory term by term,
using a tricky limiting procedure
that in modern terms amounts
to expressing the perturbation theory in terms of the
effective coupling at a small momentum typical of the
physical situation considered.
The convergence of the perturbation theory, upon
which the renormalization procedure hinged,
was very doubtful (in fact it fails to converge
for almost any non-trivial
theory, though in favorable cases
it can be rescued by Borel resummation.)

What now appears to be the most profound point was made by
Landau [\landau ].  In modern language, his point was that in a
non-asymptotically free theory the coupling instead of
decreasing logarithmically at small distances would increase,
and inevitably become large.  Thus the procedure of expanding
in a small low-energy effective coupling only hid but did
not remove the inevitable appearance of strong couplings among
the virtual quanta, which invalidate the perturbation series.
Indeed the fundamental bare coupling, which
to satisfy the requirement of locality in a theory of particles
must be fixed at infinitely small separations,
formally diverges to infinity.
If one defines the theory by a regularization or
cut-off procedure, which roughly speaking corresponds to
specifying the coupling at a small but finite distance and
letting this distance become
smaller and smaller while adjusting the
coupling accordingly, then to obtain finite results
at finite
distances the bare coupling must be taken to zero.
But doing that, of course,
leads to a trivial, non-interacting theory.  Landau's argument
that non-asymptotically free theories cannot exist
is not rigorous, because the logarithmic running of the
coupling on which it is based can only be derived at weak coupling.
(It is a fully
convincing argument that such theories
cannot be constructed {\it perturbatively}.)
Yet later work in ``destructive field theory'' has largely
vindicated Landau's intuition, and showed that many theories,
almost certainly including QED and Yukawa's pion theory, in fact
do not exist (or are trivial) despite the fact that their
perturbative expansions are non-trivial term by term.

Developments in the late 1960s and early 70s put these issues
in a new light.  The successful use of non-abelian
gauge theories to construct models for
the electro-weak interactions [\weak ], and 'tHooft's proof of
their renormalizability [\tHren ], provided a wider perspective in which
to view the earlier success of QED.  They made it seem
less plausible
that the successful use of quantum field theory in QED was a lucky
fluke.  They also raised the possibility that the unification of
electrodynamics with other interactions would cure its most severe
fundamental problem, the Landau problem just described.

On the other hand, the success of the quark-parton model [\partons ]
in describing the results of the SLAC
deep inelastic electroproduction
experiments created a rather paradoxical situation for the
theory of the strong interaction.  The quark-parton model was
based, essentially, on an intuitive but not wholly consistent
use of {\it non-interacting\/}
field theory for the supposed constituents of
strongly interacting hadronic matter.  Landau's argument
was meant to be a {\it reductio ad absurdum\/} -- showing that
the only consistent quantum field theories must be non-interacting
at short distances, and therefore trivial.  It seemed as if Nature
accepted Landau's argument, but failed to draw the obvious, absurd
conclusion!  This craziness,
together with the vulgar problem that the
quarks were never observed as individual particles, helped foster
that skepticism
both of quantum field theory and of the real existence
of quarks, which Gell-Mann expressed so eloquently.

After this prelude, let us take up our theme.

\chapter{Understanding Asymptotic Freedom in QCD}

\section{The Challenge of Understanding}

When we first discovered it [\grosswil , \pol ]
the asymptotic freedom of QCD was simply
the result of a calculation, and we had no simple physical explanation
for it.  Indeed the technology for quantizing and renormalizing gauge
theories
was new and relatively raw at the time, and we had to work hard to
convince first ourselves and then others that our result had any
physical meaning at all.   Statements about the off-shell
behavior of Green's functions involving correlations among
unobservable particles (gluons, quarks) had, and probably still do
have,
a certain air of unreality
about them, especially given that they occur in the context of a gauge
theory, where it is so easy to write down meaningless -- {\it i.e}.
non-gauge invariant -- expressions.    
Nevertheless,
the mathematical consistency of our procedures was soon generally
appreciated.  When joined with pre-existing intuitions from the 
quark-parton model [\partons ] 
which they both roughly justified and refined, these procedures produced a
simple yet concrete and quantitative
physical picture of many phenomena in strong interaction physics, an
example of which I will discuss in {\S}2.  
Yet a
simple, satisfying 
physical picture of the phenomenon of asymptotic freedom itself
emerged surprisingly late, principally in the work of Nielsen
[\nielsen ] and Hughes [\hughes ], as I shall now discuss.

\bigskip

Part of the difficulty with understanding asymptotic freedom is that
its opposite, the screening of charge, is so common and familiar. 
The screening of charge by a dielectric medium is a concept that goes
back to Faraday.
A pointlike positive charge in a medium pulls negative charges 
toward it, partially neutralizing itself and
thus exerting a weaker influence at large
distances than if it would if it were {\it in vacuo}.   Quantum field
theory introduces the innovation that even the vacuum must be regarded
as a dielectric of sorts -- a 
polarizable medium of virtual particles.  Thus the effective
coupling at large distances in the quantum theory is modified from
classical expectations.
One would perhaps be tempted to
retain the intuition that the effective charge strength would
decrease with distance, as for an ordinary dielectric.  
And indeed this sort of behavior was shown
to occur in a variety of examples by Landau and his school.
As I already mentioned, 
Landau attached tremendous significance to these results, arguing that
any finite value for the charge attached to a pointlike particle
would be completely neutralized at finite distances.  Since it is
difficult to implement relativistic invariance except for pointlike
objects\foot{Modern string theory does this, 
but also illustrates how difficult it is to do!},  
Landau concluded that 
the only truly consistent relativistic quantum field theories are
trivial, non-interacting ones.   Later systematic
work using the mature techniques of quantum field theory in a very
wide class of examples confirmed that screening is the generic 
behavior [\zee , \colegross ] if one excludes nonabelian gauge interactions.
  
\section{Antiscreening as Paramagnetism: The Importance of Spin}


It is ironic that the physical behavior to which we will ultimately
trace asymptotic freedom was well known to Landau, and central to one
of his major interests, the quantum theory of magnetism.  In a
relativistic theory we must have the relationship $\epsilon \mu = 1$
between the dielectric constant and the magnetic susceptibility.
(Indeed one can think of $\epsilon$ as the coefficient of the
electric term $E\cdot D \propto \epsilon F_{oi}F^{oi}$ and
$\mu^{-1}$ as the coefficient of the magnetic term
$B\cdot H$ $\propto \mu^{-1} F_{ij}F^{ij}$ in the action, and these
will form an invariant combination only if $\epsilon = \mu^{-1}$.)
Thus
normal electric screening behavior $\epsilon \geq 1$ is associated
with diamagnetism, $\mu \leq 1$.  However one knows that diamagnetism
is {\it not\/}
the universal response  of matter to an applied magnetic field.
Indeed it is a familiar and important result in the theory of metals
[\metals ],
that for an ideal Fermi gas of
non-interacting electrons the Landau diamagnetism associated with 
moments generated by orbital motion is
dominated by the Pauli paramagnetism arising from alignment of
elementary spin moments.  Of course this analogy is far from conclusive,
since in
metals one is dealing with a non-relativistic system of real particles
rather than a relativistic system of virtual particles.  But it does
alert one to the possible significance of spin in generating
anti-screening behavior.  Indeed we shall see that asymptotic
freedom can be quite directly attributed to the paramagnetic response
of the gluons' spin.

As a step toward making these analogies precise, it is very helpful to
re-cast the field theory problem of interest into a form resembling
a many-body problem
about which we have pre-existing experience and intuition.
The standard way to write the classical Lagrangian density for a nonabelian
gauge theory is
$$
{\cal L}~=~ 
-{1\over 4} G^a_{\alpha \beta}G^{a\alpha \beta} 
+ \bar \psi (i\gamma^\nu D_\nu -m) \psi + \phi^\dagger (-D_\nu D^\nu
-\mu^2) \phi ~ + {\rm other~ terms}  
\eqn\standardaction
$$
where the field strength is defined as
$G^a_{\alpha \beta} \equiv \partial_\alpha A^a_\beta
- \partial_\beta A^a_\alpha -gf^{abc}A^b_\alpha A^c_\beta$
with the
structure constants $f^{abc}$, and
in the matter fields the covariant derivative 
$D_\nu = \partial_\nu + ig A^a_\nu \cdot T^a$ occurs, where the
$T^a$ are the
appropriate
representation matrices (e.g., the Pauli matrices ${\sigma \over 2}$
for the fundamental representation of $SU(2)$, or the Gell-Mann
matrices ${\lambda \over 2}$ for the fundamental representation of $SU(3)$).
The `other terms' might include Yukawa couplings and scalar field
self-interactions, but do not depend upon the gauge field.
By redefining ${gA \rightarrow A}$ we can isolate the coupling
constant, in the form
$$
{\cal L}~=~ 
-{1\over 4g^2} G^a_{\alpha \beta}G^{a\alpha \beta} 
+ \bar \psi (i\gamma^\nu D_\nu -m) \psi + \phi^\dagger (-D_\nu D^\nu
-\mu^2) \phi ~ + {\rm other~ terms}  
\eqn\rescaledaction
$$
where now 
$G^a_{\alpha \beta} \equiv
\partial_\alpha A^a_\beta - \partial_\beta A^a_\alpha
                           -f^{abc}A^b_\alpha A^c_\beta$ and
$D_\nu = \partial_\nu + iA^a_\nu \cdot T^a$,
so that the only appearance of $g$ is in the coefficient of the first term.

We would like to compute the magnetic susceptibility of the vacuum,
{\it i.e}. essentially the energy density required to set up a
magnetic field of 
given strength $B$.  At first sight this is completely trivial;
from \rescaledaction\ it is just ${1\over 2g^2} B^2$.
However  \rescaledaction\ is only the classical Lagrangian,
and in interpreting
it quantum mechanically we must be careful, because the
innocent-looking classical fields become rather singular unbounded
operators ... In other words, we must do something about regulating the
short-distance ultraviolet diverges that appear when one actually computes
any specific process starting from \rescaledaction .  A pragmatic
approach well suited to our purposes is simply to throw out the modes
whose energy exceeds a cutoff value $\Lambda$.  Then when we compute
the total energy associated with the magnetic field, 
we will have to take into account zero-point energy
associated with the modes of the various vector, spinor, and scalar fields.
The value of these zero-point energies depends, of course, on the
strength of the ambient magnetic field.  It is precisely this effect
that gives non-trivial dependence, beyond the classical result, of the
vacuum energy on the value of an applied magnetic field.
This phenomenon is really quite
analogous to the corresponding situation for metals, where now the
Dirac sea is playing the role of the filled Fermi surface. 


Reserving the technical 
details of this computation for the following section, let
me now quote the answer.  One finds for the zero-point contribution 
$\Delta {\cal E}$ to
the energy density the expression
$$
{\cal E} + \Delta {\cal E} ~=~ 
{1\over 2g^2(\Lambda^2 )} B^2 ~-~ {1\over 2}\eta B^2 \ln ({\Lambda^2\over B } )
  ~+~ {\rm finite}~,
\eqn\zeropointE
$$
where
$$
\eta ~=~ {1\over 96\pi^2}
[-(T(R_0 ) -2T(R_{1\over2}) + 2T(R_1 ) )] ~+~
{1\over 96\pi^2} [3(-2T(R_{1\over 2}) +
8T(R_1)) ]  ~,
\eqn\contributions
$$
and the terms not displayed are finite as $g\rightarrow 0$ and
$\Lambda \rightarrow \infty$.  
The reason for the notation $g^2(\Lambda^2 )$ will emerge 
presently.  The factor $T(R_s )$ is
the trace of the representation for spin $s$, and basically represents
the sum over the squares of charges for the particles of that spin.
(Each species of course contributes proportional to the square of its
charge; one power for the force felt in a given field, and a second
power for the strength of field generated due to a given response.) 
The denominator of the logarithm has
been 
chosen by dimensional analysis; with this choice
one expects -- and finds -- no anomalously
large finite parts.

The different pieces of \contributions\ have a very simple and
appealing intuitive explanation, as follows. 
Imagine starting with the cutoff at $\Lambda$ and moving
it
to a smaller $\Lambda^\prime$ by integrating out the modes with
intermediate energies.  We see that these modes contribute
$$
\eqalign{
\delta ({\cal E} + \Delta {\cal E} ) ~&=~  
  -{1\over 2}\eta B^2 \ln ({\Lambda^2 \over {\Lambda^\prime}^2 }) \cr
~&=~ ({1\over \mu } - 1 ) {1\over 2g^2} B^2 ~.\cr } 
\eqn\suscept
$$
In the second line we have indicated the interpretation of the result 
as a
susceptibility from the indicated modes.  Here is the rigorous embodiment
of the heuristic ``susceptibility of the vacuum''.  

Given this now very concrete interpretation, it is easy to
appreciate the physical content of 
the formula \contributions .  For small $g$, we 
have 
$$
\eqalign{
\mu - 1 ~&=~  \eta g^2 \ln ({\Lambda^2 \over \Lambda^{\prime 2} })~ \cr
~&= ~ {g^2\over 96\pi^2} \ln ({\Lambda^2 \over \Lambda^{\prime 2} })
[-(T(R_0 ) -2T(R_{1\over2}) + 2T(R_1 ) )] \cr~&+
{g^2\over 96\pi^2}\ln ({\Lambda^2 \over \Lambda^{\prime 2} }) 
~[ 3(-2T(R_{1\over 2}) + 8T(R_1)) ]  ~,\cr} 
\eqn\explsus
$$
in which the contribution to the susceptibility from modes with
energies
between 
$\Lambda$ and $\Lambda^\prime$ is identified explicitly.  
As in the theory of metals, there are two components to the magnetic
response: the first half of \explsus ,
which arises from the orbital motion of the charged particles,
which is induced by the magnetic field and generates a counter
magnetic moment (Lenz's law); and the second half, which is induced by the
tendency of spin moments to line up with the applied field and
enhance it.  For free electrons with a gyromagnetic ratio 
$g_m = 2$, the paramagnetic response is three times as strong as the
diamagnetic response [\metals ].  The same result holds true for 
the abstract spin-${1\over 2}$ fermions in our
case -- indeed, 
basically the same calculation is involved.  For spin-1 particles in
non-abelian gauge theory, the gyromagnetic ratio is again 2, as we
shall explicitly derive below.  This value is no accident, but is
required by very general arguments about high-energy behavior [\swein ].
Since the spin for the vector bosons is twice as large,
there is a factor of four enhancement of the paramagnetic
contribution compared to fermions.  Summing over the degrees of 
freedom, we have now `derived' 
\explsus , up to an overall normalization, using nothing but
simple qualitative arguments.  The possibility of $\mu > 1$ is
already manifest in \explsus , and as we have seen this implies,
in a relativistic theory
the possibility of antiscreening behavior for electric charge.

Actually, to be perfectly
honest I should now mention explicitly two small subtleties that have
been smuggled into the equations.  One subtlety is that the
spin-1 particles must be regarded as having just two degrees of freedom
-- the longitudinal component is unphysical.
The second is that for
virtual spin-${1\over 2}$ particles the overall sign must be reversed
compared to
to response of real particles.  This is because we are computing a
zero-point energy, and the zero-point energy
of fermions is negative.

Having understood the basic physical phenomena it embodies, let us now
return to consider the implication of \zeropointE\
for observable quantities.
It is imperative,  in our description of what ought to be
a physically
meaningful, unambiguous quantity, 
to remove reference to the conventional
and arbitrary cutoff $\Lambda$.  This is a standard problem for the
renormalization program [\kramers , \schwinger ].  The
point is that if we are working with probes characterized
by energy and momentum scales well
below $\Lambda$ we expect that our capacity to affect, or be
sensitive to, the high-energy
modes will quite restricted, so that we should be able
to form a useful approximate
description in which these modes
are integrated out and do not appear as
independent dynamical variables.  In this low-energy description, we
should demand that physical quantities are essentially 
independent of the precise value of the cutoff, as long as it is
large.
We can achieve this for \zeropointE\ if we define an effective
coupling in such a way that the right-hand side becomes independent 
of $\Lambda$, as follows:
$$
{\rm const.}  ~\equiv~ 
{1\over g^2 (\Lambda^2 )} ~-~  \eta \ln ({\Lambda^2 \over B} ) ~.
\eqn\effcoup
$$
Even better, we can remove some clutter by using the differential form
$$
{d\over d (\ln \Lambda^2 )} ({1\over g^2 (\Lambda^2 )} ) ~=~ \eta~.
\eqn\diffrunning
$$
We see that the effective coupling decreases as a function of the
cutoff scale $\Lambda$, tending as the
inverse logarithm of $\Lambda$ toward zero as
$\Lambda \rightarrow \infty$, as long as there are not too many
quarks,
{\it i.e}. for $\eta > 0$.  Although we have done a
calculation that is only valid to the first order in $g^2$, the
conclusion that once the coupling becomes small it settles toward zero
is self-consistent and secure.  This behavior of the effective
coupling, its running with energy scale, is the essence of asymptotic freedom.

Now in order to get an accurate result for the energy, we should choose a
cutoff $\Lambda^2$ of order $B$.   Otherwise the logarithm in
\zeropointE\ will become large, and perturbation theory will not be applicable.
Summarizing then our calculation of the energy associated with a magnetic
field, we have
$$
{\cal E} ~=~ {1\over 2 g^2(B)} B^2~ +~ O(B^2)~,
\eqn\energyans
$$ 
where on the right-hand side quantum corrections are implemented with
a cutoff at the energy scale $\sqrt B$, and no large logarithms
occur in the corrections.  As this scale is increased --
in other words, as the cutoff is removed -- the effective coupling
$g(B)^2$ shrinks.  
Tracing back through the logic, we recognize that it is indeed the
paramagnetic response of the charged spin-1 gauge bosons present in
non-abelian gauge theories which, by ensuring
$\eta > 0$, was responsible for the smallness of the
effective coupling at large energy (or magnetic field) scales.

So far we have been discussing what might appear to be a very special and
somewhat esoteric aspect of our theory, that is the energy associated
with
a magnetic field.  It may not be transparently
obvious how to connect this to electric charge screening, or to more
general physical processes.  But actually these connections are not
very hard to see.  Indeed they are more or less immediate consequences
of the fact that, due to restrictions imposed by gauge and
relativistic invariance, there is only a single inverse coupling parameter 
${1\over g^2}$ 
appearing in \rescaledaction\ which governs the `stiffness' or
difficulty in exciting
the gauge fields.  We have found that in the quantum theory this must
be regarded as a running parameter, whose value depends on how far we
have integrated out the high-energy modes.  In a low-energy process we
will find that it is appropriate (technically, as in the discussion
around \energyans , in order that we avoid large
corrections) to integrate more of the modes out, and we will find a
larger
effective $g$; conversely, of course, for high-energy processes the
appropriate effective
$g$ becomes small\foot{
This formulation is somewhat loose, in that for
most interesting physical processes it is not possible to make a
unique clean identification of an overall large 
energy scale; but it indicates the
essence of the matter.  There is a highly developed technique, one
small portion of which we shall sample in {\S}2, for isolating
observables which do have a characteristic large scale.}.

\section{The Core Calculation}

Having argued on qualitative
grounds that its form is physically plausible,
and that asymptotic freedom is its direct consequence,
now let us actually demonstrate \contributions .

The
paramagnetic
contribution to $\eta$ from spin components $\pm s$ is easy to compute.
Let us assume unit charge and
gyromagnetic ratio $g_m$.  We are
interested in highly virtual modes, with a cutoff much larger than the
mass of the particles involved, so we will take them to be massless
(but ignore the infrared divergences this entails).  One
has the shifts in energy 
$E^2 = k_1^2 + k_2^2 + k_3^2 \rightarrow E^2 \pm g_mBs$.  Thus for the
zero-point energy density there is the shift
$$
\Delta {\cal E} ~=~ 
\int^{E=\Lambda}_0 {d^3k\over (2\pi)^3} {1\over 2}
  (\sqrt{k^2 + g_msB} + \sqrt{k^2 - g_msB} -2 \sqrt{k^2})~.
\eqn\spinshift
$$ 
Expanding to second order in $B$ and doing the angular integrals
one finds quite simply
$$
\Delta {\cal E} ~=~ 
- B^2 (g_ms)^2 {1\over 16\pi^2 } \int^{\Lambda^2}_0 {dk^2\over k^2}~.
\eqn\spineta
$$
This is essentially the second half of \zeropointE , with $g_m = 2$
and $T(R)=1$.  The group theoretic factors are easy to restore, and
will be left as an exercise for the reader.  I shall return shortly
to address the question of the numerical value of the
gyromagnetic ratio.

It is trickier to compute
the contribution to $\eta$ from orbital motion.
Let us describe our  constant magnetic field in
the Landau gauge: $A_y = Bx$ with other components vanishing.  Then
the Klein-Gordon equation reads
$$
[E^2 + {\partial^2\over \partial x^2} + ({\partial \over \partial y}
-iBx)^2  + {\partial^2\over \partial z^2} ] \phi ~=~ 0~
\eqn\kgeqn
$$
and has solutions of the type
$$
\phi~=~ e^{i(k_2y + k_3z)} \chi_n (x-{k_2\over B})
\eqn\kgsln
$$
with eigenvalue $E_n^2 = k_3^2 + B(n+ {1\over 2})$,
where $\chi_n$ is the standard harmonic oscillator wave function for an
oscillator of frequency the cyclotron frequency $\sqrt B$.  
These solutions are highly degenerate, labelled by the integer Landau
level parameter $n$ and the momentum $k_3$.  If we consider the states
within a cube of side $L$, then the displacement ${k_2/B}$ of the
oscillator must satisfy $0\leq k_2/B \leq L$, 
which shows that the density of states
in an interval $\Delta k_3$ is
$\Delta k_2 \Delta k_3 /(2\pi)^2 = 
{B\over 4\pi^2} \Delta k_3 $ per unit volume for each value of $n$.  

Thus we have for the zero-point energy density 
$$
\eqalign{
{\cal E}_0 ~&=~ 
{B\over (2\pi)^2}\sum_{n=0}^{[{\Lambda^2\over B} -{1\over 2}]} 
2\int_0^\infty dk_3 \Theta (\Lambda^2 -k_3^2 -B(n+{1\over 2})) 
{1\over 2}\sqrt {k_3^2 + B(n+{1\over 2} )}~ \cr
~&\equiv~ \sum_{n=0}^{[{\Lambda^2\over B} -{1\over 2}]} f(n+{1\over 2})~.\cr}
\eqn\diazpt
$$   
(There is a factor of 2 because $k_3$ can be either positive or negative.)
This expression is rather awkward due to the appearance of sums
rather than integrals.  For our purposes, it is sufficient to 
stop after the first non-trivial term of the Euler-Maclaurin expansion
$$
\sum_{n=0}^p g(n+{1\over 2}) ~=~ 
\int_0^{p+1} dn g(n) - {1\over 24}( g^\prime (p+1) - g^\prime (0) ) +
...
\eqn\eulermac
$$
because higher order terms will bring in extra powers of
$B/\Lambda^2$.\foot{Readers unfamiliar with the Euler-Maclaurin
formula might enjoy checking it on $g(x) = x^2$.}  You may check,
applying
\eulermac\ to \diazpt ,  that
the
integral term is independent of $B$.  The important term comes from
the derivative at zero, and takes the form
$$
\eqalign{
{1\over 24}f^\prime (0) ~&=~ 
{1\over 24} {B\over 4\pi^2}2\int^\Lambda dk_3 {B\over 2\sqrt {k_3^2}}\cr
~&=~ {B^2\over 2} {1\over 96\pi^2 } \ln \Lambda^2 ~.\cr}
\eqn\evaldiam
$$
This gives us the remaining (first) half of \zeropointE .  

There are no cross-terms between the orbital and spin pieces: the
shift in the range of integration is opposite for opposite spins, and
so the modification of the Euler-Maclaurin endpoint contribution
cancels between them.

To complete our discussion of the
core calculation, it remains only to justify the
choice $g_m=2$.  An easy and appropriate way to do this is to show
that the equation for the energy levels when the  component of spin 
along the magnetic field direction is $s$ gets
changed from \kgeqn\ to
$$
[E^2 + {\partial^2\over \partial x^2} + ({\partial \over \partial y}
-iBx)^2  + {\partial^2\over \partial z^2} + 2Bs] \phi ~=~ 0~ 
\eqn\spinE
$$
-- indeed, this is precisely the formulation we
made use of in the derivation.

For
elementary (unit charge, massless)
fermions satisfying the Dirac equation it suffices to note that
$$
\gamma^\mu D_\mu \gamma^\nu D_\nu ~=~ 
g^{\mu \nu} D_\mu D_\nu + {i\over 4}[\gamma^\mu , \gamma^\nu ] F_{\mu
\nu}~.
\eqn\spinorg
$$
Indeed the first, Klein-Gordon,  term gives \kgeqn\ and the second
term corrects it;  a trivial gamma-matrix manipulation casts this
correction 
into the
desired form \spinE\ for $s=\pm {1\over 2}$.  

For the vector bosons it
is essential to start from the proper Yang-Mills equations.  We will
obtain a description of unit charge spin-1 vector mesons by taking an
$SU(2)$ Yang-Mills theory and identifying the third component of 
internal isospin as the measure of charge.  
The Yang-Mills equation reads\foot{In this paragraph I will treat the
indices as sub- or super-scripts according to typographical convenience.}
$$
(\delta^{ab} \partial_\mu + \epsilon^{abc} A^{(c)}_\mu )
(\partial_\mu A^{(b)}_\nu - \partial_\nu A^{(b)}_\mu - \epsilon^{bde}
A^{(d)}_\mu A^{(e)}_\nu)
~=~ 0~. 
\eqn\ymeqn
$$
To get the equation for the charged field we add the $a=1$ component
of \ymeqn\ to $i$ times the $a=2$ component.  We also linearize in the
fluctuating fields $A^{(1)}, A^{(2)}$.  Then defining 
$
A^{(+)} \equiv A^{(1)} + i A^{(2)}
$
and removing the unphysical longitudinal mode by imposing
$$
D_\mu A^{(+)}_\mu ~\equiv ~ (\partial_\mu - i A^{(3)}_\mu)A^{(+)}_\mu
~=~ 0
\eqn\longmode
$$
one finds after some straightforward algebra the equation
$$
D_\mu D_\mu A^{(+)}_\nu 
+ 2i (\partial_\mu A^{(3)}_\nu - \partial_\nu A^{(3)}_\mu ) A^{(+)}_\mu 
~=~ 0
\eqn\chargedvector
$$
for $A^{(+)}$.  The first term on the left-hand side, with its
covariant derivative,
justifies the identification of $A^{(+)}$
as a unit charge field.  Putting in the circular polarization vectors
${1\over \sqrt  2} (1, \pm i , 0)^{\rm T}$ that correspond to unit
spin along the $\hat z$ direction of the magnetic field, we see that 
the second term indeed corresponds to the gyromagnetic ratio $g_m = 2$
for \spinE .

\bigskip

The calculation presented above is quite direct and elementary, but
the tricky integrals and non-covariant methodology are rather alien to
the usual technique of perturbative QCD, which uses standard Feynman
graphs and covariant perturbation theory.  Can
one identify the separate orbital and spin contributions in more
elegant ways?  Indeed one can, in at least two ways, as I shall now
briefly indicate.  

The one-loop corrections to the effective action, 
which is  -- in another
language --
what we have been concerned with, are given as logarithms of 
powers of the
determinants
of the operators in the action acting on the various fields.  For
scalar fields we get the logarithm of the inverse square root of the
determinant of the Klein-Gordon operator which acts on  the
scalar field;  for the spinor
(quark) fields we get the logarithm of the determinant of the Dirac 
operator; and  for
the vector (gluon) fields we get the logarithm of 
the inverse square root of the determinant
of the modified Klein-Gordon operator that appears in \chargedvector .
For the scalar and spinor fields, these
determinants are calculated very efficiently 
as the simple one-loop vacuum polarization graphs.  The running of the
coupling in scalar or spinor QED can be calculated this way in a few
lines,
and it was known for at least 20 years  prior to the discovery of
asymptotic freedom [\euheis ].

With a little
ingenuity and a lot of hindsight we can now see, without
calculation, that the QCD
result is essentially implicit in the QED result, as follows. 
To get the determinant of the Dirac operator for the spinor fermions
we may take the square root of the determinant of the square of this
operator, which is the Klein-Gordon operator modified by the magnetic
moment term as in \spinorg .  Now we recognize that we could
do the calculation in an alternative way by using the perturbation rules
appropriate to \spinorg , including an explicit magnetic moment
coupling and a quadratic propagator.  In this form the calculation
resembles that for scalar particles, but we must remember to change
the overall sign since we want the log of the 
determinant whereas by treating
the spinors as bosons we shall obtain the inverse determinant. (Note that
the square root has taken care of itself, since we squared the Dirac
operator.)  Thus by comparing the known spin-0 and spin-1/2 results we
can determine the coefficients of the two terms in
\contributions\ separately.  Since the form of this expression, given
those
coefficients, was derived from very general considerations, the
complete answer -- which, of course, includes asymptotic freedom -- 
follows immediately.

Another very elegant way to get the effective action is to use the
Schwinger proper time technique [\proptime ].  This technique 
allows one to find the
effective action in a constant field simply and exactly from the
single-particle dynamics; and it also naturally separates the orbital and
spin contributions.  But to discuss it properly would take too much time,
and I want now to turn to other, more down-to-earth matters.

\chapter{Asymptotic Freedom in Action: Gluonization of the Proton}

In {\S}3 I shall give a very
broad-brush sketch of some of the
outstanding phenomenological applications of asymptotic freedom; but
first I would like to discuss one, that is both topical and particularly
close to my heart, in a little more depth.

\section{`X-Raying' the Proton at Various Resolutions}


Though in a broad sense one could regard both classical microscopy and
atom-smashing experiments in the tradition of Rutherford and
Geiger-Marsden as precursors, sophisticated conscious use of
scattering experiments to probe the microscopic structure of matter
really came into its own with the development of x-ray diffraction.
The deep inelastic scattering experiments of Friedman, Kendall, and
Taylor [\fkt ],
which marked an epoch in particle physics, represent in many
respects the logical continuation of this development into the
sub-protonic domain.   The relativistic nature of the
phenomena in this domain, evidenced concretely 
in the abundant production of 
antiparticles, introduces essentially new features.  How can one
discuss the interior structure of an object, when the number and type
of its constituents is ill-defined?  What does it mean, to take a 
`snapshot' of an object, when its constituents move at close to the
speed of light and the limitations of the concept of simultaneity,
always present in principle, really start to cut?
Because of questions like these, it requires care and some special
technique to extract the important dynamical, space-time information  
contained in the appropriate experimental measurements.


In deeply inelastic scattering, one bombards the hadronic target --
in the simplest case, a proton -- with high-energy leptons, and
observes the momentum and energy of scattered leptons.  The
fundamental electromagnetic or weak interaction couplings of the
leptons are assumed known, and those of the hadrons are summarized
in the matrix elements of appropriate vector and axial currents $J_\mu$.  
Within this framework, then, the experimental procedure
can be regarded as an examination of the proton with 
well-characterized
probes.
For simplicity I shall 
mostly discuss the electromagnetic process here.

In an inclusive
measurement one sums over the final state, so the object of interest
is the current product  
$$
\eqalign{
W_{\mu \nu}(p,q) ~&=~ {1\over 4m}\int {d^4y\over 2\pi}
e^{iq\cdot y}\langle p|[J_\mu({y\over 2}) J_\nu(-{y\over 2})] |p \rangle \cr
            ~&=~ 
-(g_{\mu \nu} - {q_\mu q_\nu \over q^2}) W_1(x, Q^2) \cr~&+~
                 {1\over m^2} (p_\mu - {p\cdot q\over q^2}q_\mu)
               (p_\nu - {p\cdot q\over q^2}q_\nu) W_2(x, Q^2) \cr }
\eqn\currprod
$$  
where for simplicity I have taken the 
average over proton spin, although there is much
interest these days in the additional amplitudes that appear 
when this variable is restored.  The appearance of the commutator (as
opposed to the product), will be justified momentarily. 
Here of course $p$ is the proton's 4-momentum and $m$ its mass,  
$Q^2 \equiv -q^2$, and  $x \equiv {Q^2\over 2pq} \equiv {q^2\over 2\nu}$.
The cross-section is given in terms of these quantities as
$$
{d^2\sigma \over d\Omega dE^\prime }~=~ {\alpha^2\over 4E^2
\sin^4{\theta\over 2}} (2W_1 \sin^2 {\theta \over 2}  + W_2 \cos^2
{\theta \over 2} )~, 
\eqn\cross
$$
where $E, E^\prime$ are the energies of the initial and final lepton,
and $\theta$ the scattering angle.  Evidently, then, $W_1 (x,Q^2)$ and
$W_2(x, Q^2)$ are conveniently observable quantities.

The invariant mass$^2$ of the final state is
$$
(p+q)^2 ~=~ m_p^2 + 2pq + q^2 \equiv m_p^2 + Q^2 ({1\over x} -1 )~\geq m_p^2,
\eqn\masssq
$$
Thus changing the sign
of $q$ or equivalently $y$ on the right hand side of  \currprod\ leads
to a vanishing
quantity, and one can replace the current product by a commutator.

Most interest attaches to isolating the influence of the fundamental
degrees of freedom.  In quantum field theory, these manifest
themselves most clearly in the singularities they generate when
propagating (as virtual particles)
along approximately light-like world lines.  Thus it proves fruitful 
to
focus on the behavior in the kinematic regime where simultaneously
a large energy is transferred to the proton and 
the influence of
the $x^2 \approx 0$ region is amplified.   To identify this regime let us
go to a frame in which the proton is at rest and the three-momentum
transfer
is in the $\hat z$ direction; then
$$
q ~=~ ({\nu \over 2m}, 0,0, \sqrt { ({\nu\over 2m})^2 + Q^2 })~.
\eqn\qvector
$$
Thus for large $\nu$ we have for the so-called
light-cone variables
$$
\eqalign{
q^0 + q^3 &\sim {\nu \over m} \cr
q^0 - q^3 &\sim q^2 / {\nu\over m } = -m/2x ~. \cr }
\eqn\qlcvar
$$
The integrand in \currprod\ will therefore 
be very rapidly oscillatory  unless
$$
\eqalign{
|y^0 + y^3 | &\lsim 2x/m \cr
|y^0 - y^3 | &\lsim m/\nu ~.\cr  }
\eqn\ylcvar
$$
The restrictions of \ylcvar\ are consistent with $y^2 \approx 0$ only if
the components $y^1, y^2$ of $y$ transverse to $\vec q$ are small of order
$m/\nu$.  

We
can expect to get
the
most interesting information, relating directly
to the fundamental degrees of freedom, 
by studying the limit $\nu \rightarrow  \infty$ at fixed $x$, because
in this limit one finite space-time parameter ($y^0 + y^3$) is fixed and the
others scale uniformly to zero.  In
other words, 
these elementary considerations direct us toward
the famous Bjorken limit [\bjlim ].  

If the singularities 
of the current commutator, as a function of nearness to the light-cone at
a fixed distance from the origin, have a simple scaling form, then
so will the structure functions $W$ as functions of $Q^2$ and $x$.
This hypothesis is of course true in free field theory,
and leads to a very specific form, {\it i.e}. 
Bjorken scaling [\bjlim ].  In an interacting field theory, 
one in general anticipates deviations from this behavior.  These
deviations are in some sense minimized in an asymptotically free
theory such as QCD but, as we shall soon see dramatically emphasized,
they are by no means uniformly small.

The expression \currprod\ with a commutator is given by twice the
imaginary part of the expression with a time-ordered product.  This in
turn is related by a dispersion relation in $q^0$, {\it i.e}. $\nu$,  
taken at fixed large $Q^2$ so as to probe the Bjorken limit, 
to the amplitude governing forward ``Compton'' scattering
for the current in question -- all this being a sophisticated
version of the optical  theorem.  In this way the deep inelastic process,
treated inclusively, becomes quite analogous to coherent x-ray
diffraction as a probe of the structure of matter, 
with the crucial additional feature that the
variable $Q^2$ is adjustable -- as if one had access
to x-rays of arbitrary (actually, negative) mass$^2$.

\section{Partial Wave, or Moment, Analysis [\gwtwo , \gp ]}

For a reason that will soon become apparent,
we will want to focus our analysis around the singularities in the 
time-ordered product
of the currents at nearly light-like separations -- {\it i.e}.
precisely
the sort of product that governs high-energy virtual Compton
scattering, as
just mentioned above.  
These singularities are best exposed using  
Wilson's operator product expansion [\wope ], {\it viz}. 
$$
\eqalign{
{J_\mu}(a;{1\over2}~y){J_\nu}(b;-{1\over2}~y)&={1\over2}~g_{\mu\nu} \left(\partial
\over \partial
y \right)^2 {1\over y^2-i \epsilon y_{o}} \sum_{n=0}^\infty \sum_{i}
C_{i,1}^{(n)} (a, b; y^2 -i \epsilon y_{o}){\cal O}^i_{\mu_1}\cdots_{\mu_n}
(0)y^{\mu_1}\cdots y^{\mu_n} \cr
&+{1\over y^2-i \epsilon y_{o}} \sum_{n=0}^\infty
\sum_{i} C_{i,2}^{(n)} (a, b; y^2 -i \epsilon
y_{o}){\cal O}^i_{\mu \nu \mu_1}\cdots_{\mu_n}(0)y^{\mu_1}\cdots
y^{\mu_n} + ... 
~,\cr}
\eqn\wilsonop
$$
where the omitted terms are subleading for $y^2 \rightarrow 0$.  The 
${\cal O}$ represent local operators with the appropriate
quantum
numbers.  From the form of \wilsonop\ , one sees immediately
that they may be taken as
traceless, and that the most important operators of each spin are
those
of the lowest mass dimension, as these will tend to have the most
singular coefficients.  In principle the degree of singularity of
the
coefficient functions is independent of the mass dimension of the
operators -- anomalous dimensions -- so that the formally subleading
operators could in reality dominate, but in an asymptotically free
theory, as we shall see, corrections to naive scaling are only logarithmic.    
The indices $a,b$ on the currents are flavor indices, meant to convey
the possibility 
to probe different combinations of operators using various
weak\foot{For weak, 
parity-violating currents there is an additional important
structure function
$F_3$ in \currprod .}  
electromagnetic currents and scattering off different nuclear targets.

In QCD the relevant leading operators are easily identified to be
$$
\eqalign{
^{n}{\cal O}_{\mu_1}^{V} \cdots _{\mu_n} &= i\ ^{n-2} S\ {\rm Tr} \
G_{\mu_{1} \alpha} \nabla_{\mu_2} \cdots \nabla_{\mu_{n-1}} G^{\alpha}
_{\mu_n} \cr
&{\rm -trace ~ terms,} \cr
}
\eqn\glueop
$$

$$\eqalign{
^{n}{\cal O}_{\mu_1}^{F , s} \cdots _{\mu_n} &= \ i\ ^{n-1} S\
\overline{\psi} \gamma_{\mu_1} \nabla_{\mu_2} \cdots \nabla_{\mu_n} 
\ \psi \cr
&{\rm -trace ~ terms,} \cr
}
\eqn\fsingop
$$
$$\eqalign{
^{n}{\cal O}_{\mu_1}^{F , a} \cdots _{\mu_n} &= \ i\ ^{n-1} S\
\overline{\psi} \gamma_{\mu_1} \nabla_{\mu_2} \cdots \nabla_{\mu_n} 
{1 \over 2}\ \lambda^{a} \psi \cr
&{\rm -trace ~ terms,} \cr
}
\eqn\fnsingop
$$
where the color indices have been suppressed, and \fsingop\ , 
\fnsingop\ are respectively the flavor singlet and non-singlet operators.
$S$ denotes symmetrization over the Lorentz indices.

The Wilson coefficients are directly related to observables according
to 
the moment relations
$$\eqalign{
\int\nolimits^{1}_{0} dx \ x^{n} F_{1}^{a,b} (x,Q^2) &= \sum_{i} 
{\tilde C}_{i,1}^{(n+1)}  \ (a, b; Q^{2}) \ M_{i}^{n+1} ~,\cr
\int\nolimits^{1}_{0} dx \ x^{n} F_{2}^{a,b} (x,Q^2) &=  \sum_{i} 
{\tilde C}_{i,2}^{(n+2)}  \ (a, b; Q^{2}) \ M_{i}^{n+2} ~,\cr
}
\eqn\momentrels
$$
where following a universal though somewhat redundant
convention we introduce
$F_1 \equiv W_1$, $F_2 \equiv {\nu\over m} W_2$, and where
$$\eqalign{
{\tilde C}_{i,k}^{(n)}  \ (a, b; Q^{2})&={-i\over 2} \ ({Q^{2})}^{n+1} \left
(-{\partial \over \partial Q^{2}}\right)^{n}\cr
&\times \int d \ ^{4} y e ^{i q \cdot y} {C_{i,k}^{(n)}  \ (a, b;
y^{2}) \over y^{2} - i \epsilon y_0}~,\cr
}$$
and
$$\eqalign{
\langle p\vert O^{i}_{\mu_1}\cdots _{\mu_n} (0)\vert p \rangle & _{\rm {spin~
average}} \cr
&=i^{n} {1\over m} p_{\mu_1}\cdots p_{\mu_n} M_{n}^{i}\cdots ~,
\cr}
\eqn\matrixelts
$$
up to terms suppressed by $1/Q^2$.
These relations are derived by substituting the operator product
expansion
for the time-ordered product into the dispersion relation connecting
this product 
to the commutator, and identifying coefficients in the high-energy
asymptotics at fixed $Q^2$.
This somewhat 
roundabout procedure is necessary because
the expansion of the commutator cannot be used
directly; indeed,
when inserted into integrals like \currprod\
it is ordered in powers of 
${pq\over q^2} \sim {1\over x}$, and only converges for very large
$x$, which is the unphysical region.

The coefficient functions in the operator product expansion are
both fundamental and convenient, because they relate directly to 
the operator structure of the theory, without reference to the
very complicated eigenstate structure.  One consequence is that
they obey simple renormalization group equations of the form:
$$
Q^2{\partial \over \partial Q^2} \tilde {{C}}^{(n)}(Q^2, g)~=~ 
(\beta (g) {\partial \over \partial g} - \gamma^{(n)}) 
\tilde {{ C}}^{(n)}(Q^2, g)~.
\eqn\rgeqn
$$
Here of course 
$$
Q^2 {dg(Q^2)\over dQ^2} ~\equiv~ \beta (g) 
\eqn\runbeta
$$
encodes the running of the coupling as discussed in the {\S}1,
and the $\gamma$s are the so-called
anomalous dimensions of the indicated
operators.
These equations have a very transparent physical interpretation, along
the lines of our earlier discussion of the running coupling.  In that
discussion, we saw that it is appropriate to renormalize the charge
as one considers physical processes at different momentum scales.
Formally, this was embodied in the process of renormalizing the 
operator $G_{\alpha \beta}^aG^{a\alpha \beta}$ that appears in the 
Lagrangian.  For the same reasons, it is appropriate to renormalize
the operators that appear in the operator product expansion; and this
is the physics embodied in \rgeqn .  Note that when one has two
operators
with the same quantum numbers they generally mix under
renormalization; thus the gluon operators \glueop\ and the singlet quark
operators \fsingop\ are coupled together by
a non-trivial 2$\times$2 matrix $\gamma$.

These equations can be used 
to relate the coefficient functions at one $Q^2$
to those at another, by the method of characteristics.  
Indeed, in the flavor non-singlet case, where the equation has
only one component, we have
$$
\eqalign{
&{\tilde C}^{(n)} (Q^2, g(Q_0^2)) ~=~ \cr
&\exp \bigl( -\int^{\ln Q^2}_{\ln Q_0^2}\gamma^{(n)}(g(Q^2))~ d(\ln Q^2)\bigr)
{\tilde C}^{(n)} (Q_0^2 , g(Q^2) ) ~.\cr}
\eqn\coeffev
$$
In the flavor-singlet (two-component) case we have a similar
equation with ${\tilde C}$ interpreted as a two-component vector and
an ordered matrix exponential is understood in \coeffev . 

Because of asymptotic freedom, the coupling $g(Q^2)$ gets driven to 
small values at large $Q^2$, and one exploits \coeffev\ together with
perturbation
theory to obtain approximate expressions for the coefficient
functions,
which according to  
\momentrels\ translate directly into predictions   
for experimental observables.

\section{Gluonization of the Proton}

The formal framework in which deep inelastic scattering is properly
analyzed has now been set out; it remains to extract specific concrete
results, and to interpret them [\gwone ].

The solution \coeffev\ of  
the renormalization group equations for 
evolution of the
coefficient functions can be taken to a numerical level using
perturbation theory if the effective coupling $g(Q^2)$ 
is small over the range of extrapolation.
If on the right-hand side
one simply ignored the multiplicative prefactor -- {\it i.e}. set
the anomalous dimensions to zero -- and put $g(Q^2) = 0 $ in 
the rescaled coefficient factor, one would reproduce the results of
a free field theory of quarks: the parton model.  
There are two
types of corrections to the parton model predicted: 
multiplicative corrections,
from exponential involving
the anomalous dimensions, and additive corrections, from
expanding
the coefficient functions in the small parameter $g(Q^2)$.  The
additive
corrections are interesting, especially where they correct the
parton-model
sum rules or when (as in the correction to the Callan-Gross [\cg ] relation)
the leading term vanishes.  However the potentially large, cumulative
corrections are the multiplicative ones, and I 
would
like here to focus on them.

The numerical values of the anomalous dimensions, to lowest
non-trivial order, were first calculated in [\gwtwo , \gp].  They are
as follows:
$$
\eqalign{
\gamma^{(n)V}_V ~&=~ {g^2\over 8\pi^2} 
 \bigl[ 3 \bigl( {1\over 3} - {4\over n(n-1)} - {4\over (n+1)(n+2)} 
 + 4\sum^n_2 {1\over j} \bigr) + {2\over 3} n_q \bigr] \cr
\gamma^{(n)F}_F ~&=~ {g^2\over 8\pi^2} \bigl[ {4\over 3} 
 (1 - {2\over n(n+1)} + 4\sum^n_2 {1\over j} ) \bigr] \cr
\gamma^{(n)V}_F ~&=~ {g^2\over 8\pi^2} \bigl[ {-2n_q(n^2 + n +2 ) \over
 n(n+1)(n+2)} \bigr] \cr
\gamma^{(n)F}_V ~&=~ {g^2\over 8\pi^2} 
\bigl[ -{8\over 3} {n^2 + n+ 2 \over n(n^2 -1)} \bigr] ~,\cr }
\eqn\anomdim
$$
where the indices describe the mixing in the flavor-singlet case,
{\it e.g}. 
$\gamma^{(n)V}_F$ parametrizes how \fsingop\ mixes into \glueop\ as
the scale $Q^2$ increases, 
and $n_q$ is the number of relevant quark species.  
For the flavor non-singlet channels, only $\gamma^{(n)F}_F$ is relevant.
The general tendency is that the eigenvalues are positive and increase
with $n$, which in turn indicates that the higher moments of the
structure
functions tend to fall ever more rapidly with increasing $Q^2$.  On
the
other hand for the singlet channel and $n=2$ the anomalous dimension
matrix 
has a zero eigenvalue, corresponding to the non-renormalization of the
energy-momentum tensor.  So there is a non-trivial portion of the
area under the curve $F_2(x, Q^2)$ which is conserved as 
$Q^2 \rightarrow \infty$.  Putting this observation together with the 
previous one, we are led to expect roughly speaking that this
structure function will evolve toward something resembling a Dirac
delta function at $x=0$ as $Q^2 \rightarrow \infty$.  

It is, naturally, of the highest interest to characterize this
phenomenon
more precisely.  To do it, several of us [\sixman , \zwt ], 
within months of the
original discovery of asymptotic freedom, 
exploited Mellin transform technique both to invert the moment
problem and to extract the $x\rightarrow 0 $ asymptotics.  The Mellin 
transform is of course the standard mathematical 
tool for inverting moment problems of
the type confronting us here.  Indeed, this technique 
has two special advantages
in the present context.  First: 
its convolution theorem immediately indicates that
the factorization of matrix elements times coefficient functions on
the
right-hand side of \momentrels\ translates into a simple convolution
integral
expressing the structure function at one $Q^2$ in terms of the
structure function at some reference $Q_0^2$ and a kernel which is 
entirely determined by the evolution of the coefficient functions,
and thus by the fundamental microphysics, untangled from
any dependence on the target.  Second: it gives us an explicit
analytic expression for this kernel, in a form most suitable for 
asymptotic estimates.

This is not the place to enter into algebraic details, and I will now
simply describe the main result.  The most singular part of the kernel
comes from the right-most singularity of the anomalous dimensions,
which occurs at $n=1$ in the singlet channels that generate gluons.
This singularity in the evolution kernel evolves a structure function
which grows as $x\rightarrow 0$ in a calculable way 
from one that does not, and under
rather mild
assumptions this effect dominates the observable small-$x$ structure
functions at large $Q^2$.  It predicts the asymptotic form
$$
\ln F_2 ~\rightarrow~ 
2\bigl( {4\over 3} \ln (K \ln {Q^2\over Q_0^2 })\ln {1\over x}\bigr)^{1\over 2}
~+~ {\rm less~ singular}~,      
\eqn\asyresult
$$
where $K > 1$ is a numerical constant.  

It is quite remarkable, I think, to find such a precisely defined
but peculiar asymptotic behavior,
slower than any power but faster than any power of a logarithm,
appearing
in fundamental physical theory.  

\FIG\xzero{Comparison of the predicted behavior of the structure
function
at small $x$ with observations at HERA, from [\bffig ], to which you are
referred for details.  As discussed in the text,
the observed steep rise in the structure functions at
small
$x$ and large $Q^2$, displayed here, is a long-standing, no-parameter
prediction
of asymptotic freedom in QCD.  Plotted is a comparison of data to a
scaling curve based on (2.17). }

These predictions of 1974 waited almost 20 years before they could
be decisively compared with experiment.  Perhaps the most profound
achievement of the HERA accelerator has been precisely to accomplish
this.  The results, displayed in Figure \xzero , are extremely gratifying.
Allow me to emphasize that here we are not speaking of small
corrections,
but of wholesale ($\gsim 100$\%) violations of naive Bjorken scaling. 
The point is that effects which are perturbative, and therefore small and
calculable, within each differential  $Q^2$ interval, accumulate 
in these multiplicative corrections to build
up 
a large yet still precisely calculable total. 

One might well be concerned that perturbative calculation of the
anomalous dimensions might break down near the singularity, and 
that this would
invalidate the results we extracted by taking its perturbative form
at face value.  And indeed, if one were 
to attempt to go strictly to $x=0$ at any fixed $Q^2$, however large,
one would certainly run into trouble.  
Fortunately, R. Ball and S. Forte [\bfcalc ] have demonstrated that
one can reorganize the calculation, taking $x$ small and $Q^2$ large
in a correlated way, so that perturbation theory is manifestly valid
throughout the whole range of HERA measurements.  
So the truly remarkable fit between theory and experiment demonstrated
in Figure \xzero\ is hardly a fluke.  Indeed, by further refining the
analysis [\bfcalc ]
these authors have extracted from the
rise in the structure function with $Q^2$ at small $x$
a numerical value of the effective
coupling that is consistent in value with, and comparable in accuracy to,
the other best determinations.

\section{Interpretation}

The preceding analysis is precise and physically rigorous, but perhaps
lacking in intuitive appeal.
There is an alternative 
approach to these results, which is less rigorous but perhaps more
transparent.   It was made popular in the context of QCD by 
Altarelli and Parisi [\altpar ], 
though related discussions date back to
the
earliest days of parton theory [\partons , \ks ].

This is not the place for a self-contained review of the parton model.
Let me only recall, that in this model the deep inelastic scattering
is
modelled as scattering off point-like constituents -- originally
abstract partons, later taken to be quarks, and, of
course, in the modern formulation colored quarks and gluons.  In this
formulation, the structure functions are essentially suitably weighted
densities of these constituents, with the structure function at a
given
$x$ measuring the densities with momentum fraction $xp$ of the proton
momentum
in the infinite momentum frame.  In the original parton model these
densities
were supposed to be independent of $Q^2$; we are concerned with the
refinement of
that
picture, to take into account the internal structure of quarks and gluons.

According to this refinement, when one ``looks at'' the proton
with photons of larger and larger 
$Q^2$, one is viewing its constituents with finer and finer resolution,
and better
perceiving their internal structure.  One may find that what appeared
to be a quark 
is resolved instead as a quark + gluon,  
for example.
The constituents of the quark will have smaller momentum fractions,
{\it i.e}. smaller values of $x$.  An alternative to describe 
the same phenomenon, is to say that in response to a more violent
kick the quark has some chance of radiating a gluon.
In any case, as one takes $Q^2$ ever
higher,
the structure functions is expected to shift toward smaller and
smaller
$x$ values.  This is, of course, just the behavior we inferred from
the
more rigorous arguments and calculation.   
 
The dominant effect is that quarks and gluons bremsstrahlung soft
gluons, which in their turn bremsstrahlung soft gluons, ... .  This
simple but fundamental qualitative effect is quite directly the cause 
of the dramatic HERA phenomenon.  
It is also, at least metaphorically, the central phenomenon of
asymptotic freedom itself: when we closely examine a color charge, we
find that at its core it dissolves into a cloud of soft glue.

\chapter{The Significance of Asymptotic Freedom in Fundamental Physics}

The consequences of the
asymptotic freedom of QCD ramify throughout fundamental physics.  I
would like now very briefly to summarize the principal ones.

\section{Analysis of Data: from Tests to Backgrounds}


A great wealth of data has been
explained quantitatively by perturbative
calculation of processes characterized by a large energy and momentum
scale.  Of course, it is only because of asymptotic freedom that such
calculations make sense.  Very extensive reviews are available
[\datareviews ]; here I will only mention a few highlights:

$\bullet$ {\it Inclusive jet processes}:


\FIG\fincljet{Inclusive jet cross sections as measured by the CDF
collaboration, compared with theoretical expectations based on
perturbative QCD.  Note the scales.  More recent work has shown
improved the precision 
of the measurements and the quality of the agreement at the largest energies.}

Perhaps the most awesome verification of asymptotic freedom and perturbative
QCD, considering the energetic sweep of the experiment and the variation
in the predicted quantities over many orders of magnitude, is the one
displayed in
Figure
\fincljet\ [\incljet ].  It shows a comparison between
the cross-section for inclusive jet
production as a function of transverse energy as predicted and measured.
The jets, of course, are interpreted as indicating the energy and
momentum associated with an underlying quark or gluon that has been
produced in a high momentum-transfer collision.  One would
not expect the process of
evolution from an underlying quark or gluon into hadrons
to distort the conserved quantities of total energy and momentum in
a gross manner,
because the dressing process is soft -- the quarks and gluons interact
only weakly with highly virtual particles!  Thus, as is
borne out by more precise, detailed calculations,
the microscopic process is accurately
reflected in these observables.  At a slightly more
technical level: by taking the
inclusive cross-section, one removes sensitivity to other scales;
then, because the
underlying hard collision that produces the jet-initiating
quark or gluon involves a large energy and momentum transfer
it is governed by a small effective coupling, and accurately calculable.

$\bullet$ {\it Antenna patterns}:


\FIG\fantenna{Distribution in energy and angle of jets in 
3-jet events from Z decay, compared to QCD 
and to various straw-man competitors.  This is very direct and
tangible evidence for the existence of gluons with the correct
spin and form of coupling.}

$Z$ boson decays provide, from the point of view of QCD, 
a copious source of very clean initial states.  The lowest order
description is that the $Z$ decays into various $\bar q q$ pairs;
these are produced at large energy and relative momentum 
and thus are predicted to produce two independent jets.  At next
order in the small coupling $g^2(M_Z^2)$ there 
is the possibility of gluon emission, which produces three-jet events.
The relative frequency of such events, as well as the
detailed ``antenna pattern'' of energy and angle distribution is fully
calculable, and reflects the existence of the gluon and 
the basic underlying microscopic coupling
ultimately responsible 
for the strong interaction in
almost ideally direct form.  The measurements agree quite accurately
with the detailed predictions, as displayed in Figure \fantenna\ [\antenna ]. 

$\bullet$ {\it Heavy quark spectroscopy}:


\FIG\fheavy{Comparison of the $\Upsilon$ spectrum with calculations
based on direct
numerical simulation of QCD using a fine-grained lattice grid.
Note that the only parameters in these calculations are the mass of
the quark and the strength of the QCD coupling, both of which have
precise meaning in the underlying microscopic theory.  }

In forming the ground state of a Coulombically bound quark-antiquark
system, the most important distances are of order the Bohr radius 
$(\alpha_s m_q )^{-1}$ or less; the relevant part of the potential 
therefore derives
from exchanges at momentum transfers of order $\alpha_s m_q$ or
greater.  If this is sufficiently large that the appropriate value of
the running coupling $\alpha_s ((\alpha_s m_q)^2)$ is small, the
perturbative treatment is self-consistent and not grossly inaccurate.  
One therefore expects a spectrum which roughly resembles that of
positronium.
This expectation is famously borne out in the $\bar c c$ 
$J/\psi$ family and the $\bar b b$ $\Upsilon$ family of resonances.
When perturbation theory is cunningly combined with non-perturbative
information from lattice gauge theory, it becomes possible to
calculate the spectrum with astonishing accuracy.  Figure \fheavy\
[\heavy ]
indicates the state of the art.

$\bullet$ {\it The running coupling}:


\FIG\frunning{The coupling constant as determined from experiments
with different characteristic scales, compared with the prediction of
asymptotic freedom in  QCD.  There is no free parameter in this
comparison.  Moreover, within each experiment there are many phenomena
which the theory must, and does, describe consistently, still with no
free parameter.}

It is, of course, of fundamental interest directly to check 
that the measured coupling runs according to the predictions
of asymptotic freedom.  A plot incorporating data from a wide
variety of experiments is given as Figure \frunning ; clearly there is
impressive agreement between theory and experiment.  I will not
attempt
to do justice to the different experiments, but confine myself to
a few general remarks.

Because of the way that the strong coupling runs, decreasing as 
the (energy or momentum)$^2$ 
scale $Q^2$ increases, testing QCD by measuring the coupling as a
function of $Q^2$ has a two-faced character.  At large $Q^2$ the
predictions are very precise -- the coupling gets small, and we can do
accurate calculations.  Also mass corrections, which are
non-perturbative and generally difficult to control theoretically, are
suppressed by powers of mass$^2$ over $Q^2$.  At small $Q^2$ the
theory is much harder to control and make precise, but if you are
interested in quantitative results for $\alpha_s$ there is a large
premium for working at small $Q^2$.    

So if your goal is simply to check that QCD is right, then you want
a unique prediction and the high energy processes are particularly
favorable.  But if you want to determine $\alpha_s$ precisely
quantitatively, then the low energy determinations have a big
advantage.  This is the two-faced character I mentioned. 

These features are clearly evident in the classic plot of running
couplings measured at different $Q^2$, Figure \frunning\ [\running ].
You see that at low $Q^2$ there is a big spread in the predictions for
different values of the coupling, so
that is where you can measure what $\alpha_s$ is.

On the other hand
for very large $Q^2$ we find -- most remarkably -- a more or less
unique prediction for the value of the physical parameter
$\alpha_s(Q^2)$.
Almost any reasonable value of $\Lambda_{\rm QCD}$, the value where
formally
$\alpha_s \sim 1$ which sets the overall scale for the strong
interaction, will give you within about 10\% the same result for 
$\alpha_s$ at large $Q^2$.  
This profound result
essentially realizes, for the strong interaction,
the analogue of
Pauli's dream of calculating the fine structure constant.

\bigskip

I believe there is no longer a serious question of testing QCD as
such.  QCD has won a secure and permanent place as a valid description
of a large domain of experience.

This situation is reflected in
the phrase `QCD backgrounds' that one  now
so often hears in analyses of
high-energy experiments.  Implicit in this phrase is
the philosophy that
a deviation from QCD predictions will be treated as an additional
piece of new physics, for example as indicating that a new class of
particles or interactions exists, rather than as an indication that
QCD itself must be modified.

A profound justification for this attitude is that QCD is a {\it
principled\/} theory, based on concepts of gauge invariance and
universality that cannot be fudged.  Thus if the theory
agrees with experiment
even approximately (as, obviously, it does),
there is a powerful sense in which it -- that is,
the principles which it embodies -- must be a true, permanent insight
into Nature.

So much for the direct
experimental significance of asymptotic freedom in QCD.
Now I would like briefly to discuss how this result either
directly precipitated or helped to catalyze three conceptual
revolutions.

\section{First Conceptual
Revolution: Quantum Field Theory is Incarnated}

The discovery of QCD and asymptotic freedom
changed the way
people regard quantum field theory.  It made it clear that
{\it one must take quantum field theory, including its ultraviolet
problems and its non-perturbative aspects, deadly seriously.}
As I mentioned before, this attitude was by no means universally
shared prior to their discovery.

As I trust is by now abundantly clear,
the old ``problem'' of infinities in perturbation theory, and
the effects of highly virtual particles which give rise to them,
lie close to
the very root of all the marvelous success we have had in predicting
experimental results using asymptotic freedom in QCD.
Whereas in QED and
electroweak theory higher order effects of quantum
field theory generally
provide small corrections,
in QCD they are much larger quantitatively.
Two- and even three-loop calculations are needed to
address the data adequately.  And of course the logarithmic
infinities due to highly virtual quanta of large invariant mass
are directly responsible
for the running of
the coupling, which is the conceptual foundation of all QCD
phenomenology.
Thus to the extent that the predictions of QCD perturbation
theory are verified, the detailed structure
of quantum field theory and
its renormalization program stand dramatically vindicated.
The question is no longer the validity of these concepts, but
how to do them justice.

\bigskip

For the phenomenological success of QCD and asymptotic freedom
in describing a wide variety of hard processes using souped-up
methods of perturbation theory does not, of course,
remove the challenge of understanding non-perturbative aspects of
the theory.  Perhaps nothing exhibits this challenge so clearly
as the fundamental formula of dimensional transmutation.  QCD
at the classical level contains only a
dimensionless coupling and is scale invariant \foot{Strictly
speaking this holds only for massless quarks, but the essence
of the following argument is valid generally.}
On the other hand physical hadrons have definite,
non-zero masses.
How does a
parameter with dimensions of mass emerge from a fundamentally
massless theory? It happens because the running of the coupling,
which is an inevitable result of quantizing the theory,
implicitly defines a mass scale:
$$
\Lambda ~=~ \lim_{Q^2 \rightarrow \infty} Q e^{-c_1/\bar g^2(Q^2 )}
   [g^2 (Q^2 ) ]^{c_2}~.
\eqn\runcoup
$$
Here $c_1$ and $c_2$ are definite numbers that can be read off
from the first two terms in the
renormalization group $\beta$ function.  The main
point is that the limit
on the right-hand side exists and defines a finite mass.  All
other masses in the theory, including the masses of particles in
the spectrum, can be expressed as {\it pure numbers\/} times this
one.  Once the boundary condition for the running coupling
is determined the theory is completely fixed, there is no other
remaining parameter nor any
independent scale.  For our present discussion the
most significant
point is that all hadronic masses therefore will be, like
$\Lambda$, non-perturbative in $g^2$.  The challenge could
not be clearer.

The challenge of understanding non-perturbative effects in QCD
has led to several remarkable developments.
Perhaps the most important single result is that there is
now a convincing
case that the microscopic theory of QCD actually does give rise
to the confinement of
quarks and gluons inside hadrons.
One demonstrates this by showing,
using computer simulation,
that there is no qualitative change (phase transition) between
strong-coupling expansion of the discretized lattice
theory, in which
confinement is manifest but Lorentz invariance is violated --
and the
continuum limit, in which
Lorentz invariance is manifest but confinement
is not [\creutz ].  The application of
Monte Carlo methods, semiclassical approximations,
and large $N$ expansions [\largeN ] for quantum field theories
have been highly developed,
with QCD as one of the important original
motivations but now ramifying into many other areas.

A particularly
striking discovery
is the possibility of non-perturbative P and T violation in
QCD: the famous $\theta$ term [\tHooft ].  What seems to be the
most satisfactory approach to
understanding why this potential source of P and T violation is
in fact highly suppressed (as shown by the smallness of the
neutron's electric dipole moment) was suggested by Peccei and
Quinn [\pq ].  It involves the existence of a new light boson, the
{\it axion}, with remarkable properties [\axion ].
If axions do exist, they
may be very important for cosmology, plausibly even supplying the
astronomer's
``missing mass'', which is about 90\% of the Universe by weight.

\bigskip

The development of QCD had a curious effect on string theory.
Its immediate impact was certainly to
kill much of the interest in string theory,
which of course was originally developed as a model
of the strong interaction.  By providing a correct microscopic theory
of the strong interaction based on quite different principles --
and incorporating in a central place
point-like interactions at short
distances that are quite difficult to reproduce in a theory
containing only extended objects -- QCD removed from string theory
its initial source of motivation.  For the longer term however the
story is more complicated, and its conclusion is not yet clear.
By emphasizing that the short-distance properties of quantum field
theory must be taken deadly seriously, and that the ``problems''
encountered in perturbation theory are not
mere mathematical artifacts but rather signify deep properties of
the full theory, the development of QCD made the corresponding --
apparently intractable -- problems encountered in the perturbative
expansion of Einstein gravity seem that much more weighty.  Thus
the discovery that string theories can incorporate Einstein gravity
while avoiding its bad short-distance behavior is properly
regarded as
a powerful argument in favor of these theories.

\section{Second Conceptual Revolution: Unification Becomes a Scientific
Enterprise}

To achieve a unified description of apparently vastly different aspects
of Nature is certainly a major esthetic goal of the physicist's quest.
In the past it has also been a fruitful source of essentially new
insight: Maxwell's fusion of electricity and magnetism
transformed our understanding of optics, and vastly generalized it;
Einstein's fusion of special relativity with gravitation transformed
our understanding of space-time and cosmology.

The development of QCD and asymptotic freedom has enabled us
to add a major new chapter to the story of unification.  There
are two aspects to its contribution.  First,
{\it the mathematical resemblance of QCD to the gauge theories
of weak and electromagnetic interactions immediately suggests the
possibility of a larger gauge theory encompassing them all}.
Georgi and Glashow [\gg ]
constructed a compelling model of this kind
almost before the ink was dry on asymptotic freedom.
Second, the running of couplings removes the major
obvious -- superficial -- difficulty in the way of implementing such
an extended gauge symmetry, that is the disparity of coupling strengths
as observed at accessible energies.  Georgi, Quinn, and Weinberg
[\gqw ]
showed how to use the renormalization group as a quantitative tool in
investigating unification.
{\it The running of the couplings makes it possible
to study ambitious unification schemes quantitatively, and
compare them to observations.}

The logic that enables one to
connect unification ideas
quantitatively with low-energy observations
is as follows [\dim ].  One observes three {\it a priori\/}
independent couplings, corresponding
to the three gauge groups $SU(3)\times SU(2)\times U(1)$ of
the standard model, at low energies.  In a unified theory
these couplings are in reality not independent, but derive
from a single coupling.  The difference between their
observed values at low
energies must be ascribed to the different evolution of
the respective running couplings down from the energy
scale of unification.  The running of these couplings
is basically determined by the particle content of the
theory, given two inputs: the energy at which the large
gauge symmetry broke (often called the GUT scale), and the
value of the coupling at that scale.  Since therefore three
observed parameters arise from two input parameters, they
are {\it overconstrained}.  Given a specific unified
model, the
constraint may or may not be met.
If it is not met, we must discard the model.  If it is
met, then
that fact is
a highly non-trivial success for the model and
for the
assumptions that go into the calculation.

In connection with unification it is profoundly
important that the couplings run slowly;
that is, logarithmically
with energy scale.
Since there is a big discrepancy between the
effective strong
and weak couplings at presently observed energies,
there are factors
of the type $e^{\kappa\over \alpha}$ relating current
accessible scales to the unification scale.  In typical
models the GUT scale turns out to be of order
$10^{15}-10^{17}$ GeV.  This mass sets the scale for
exotic processes that occur through exchange of gauge
bosons which are in the unified group but not in
$SU(3)\times SU(2)\times U(1)$, including proton decay.
Also its large value is definitely smaller than, but
not incommensurate with, the
Planck energy $M_{\rm Pl.}\approx 10^{19}$ GeV.
where the gravitational
interaction becomes strong.  This closeness hints at
an organic connection between gravitation
and traditional particle physics.  On the other
hand the fact that
the GUT scale is significantly smaller than
$M_{\rm Pl.}$ makes it plausible that we can calculate
the running of the couplings all the
way to unification without encountering significant
corrections from quantum gravity.

Until a few years ago the minimal
unified model, based on the unifying gauge group
$SU(5)$, gave an adequate fit to the data.  That is,
the observed couplings satisfied, within their quoted
uncertainties, the constraint derived in the manner
described
above for this model.
This represents a truly extraordinary triumph for
quantum field theory, extrapolated far far beyond
the domain of phenomena it was designed to describe.
It also might seem at first sight to be rather depressing,
since it suggests a vast ``desert'' between present
energies and the GUT scale.  To be more precise: if we
do not believe the success of this calculation to be an
accident, we must not only take unification seriously, but
also make sure that unification schemes more elaborate
than
the simplest possible one
manage to give something close to the same answer.

Those intent on populating the desert
-- or (techni-)colorizing it -- should be required to
submit an appropriate environmental impact statement!

\FIG\sufiverun{Extrapolation of the couplings measured at low
energies towards higher energies, using the formulae for their
evolution discussed above, and assuming the minimal particle content
of the Standard Model.  The accuracy of the experimental
determinations 
are indicated by the thickness of the lines.}  

Recent, beautifully accurate
measurements of standard model
parameters from LEP and elsewhere have made it clear
that actually minimal $SU(5)$ doesn't quite work.  The
observed couplings are close to satisfying its constraint,
but the discrepancy is now well outside the error bars, as shown in
Figure \sufiverun .

There are various possibilities for addressing the
discrepancy, among which one seems especially noteworthy.
The noteworthy possibility is that the quantitative study
of unification of couplings has uncovered evidence for
virtual supersymmetry.

There is a standard litany of the virtues of supersymmetry,
probably familiar to all of you: it enables a new level of
unification, between particles of different spin; it
ameliorates the gauge hierarchy problem; it
is necessary to eliminate tachyons in superstring theory.
But in any accounting of the
virtues of supersymmetry, one entry is conspicuously
meager: the list of its experimentally verified consequences.

How does supersymmetry affect the running of the couplings?
It might seem at first glance that its effect is bound to
be catastrophic, since it roughly doubles the particle spectrum.
It might seem that all these new virtual particles would
inevitably induce a drastic change from the
nearly successful results
for minimal $SU(5)$.
However, it is an important fact that adding {\it complete\/}
$SU(5)$ {\it multiplets\/} to the theory affects the calculation
of the constraint among observed couplings arising
from unification of the couplings only very little.
This is because, roughly speaking, virtual particles forming
such complete
multiplets affects all three couplings in the same way.
They change the value of the
unified coupling
at the GUT scale, and can slightly modify the size of that
scale, but to a good approximation they leave the constraint
among observed couplings unchanged.

Since supersymmetry is basically a space-time, as opposed to
an internal, symmetry, the minimal
extension of the unification scheme
to incorporate supersymmetry does not fundamentally change
its group-theoretic structure.  
One simply doubles all the complete
$SU(5)$ multiplets that were in the original model, by adding their
supersymmetric partners.  The gluinos do not occur in
complete multiplets, but their contribution has the
same structure as that of the ordinary gluons,
and therefore they do not alter the
group-theoretic
structure of the calculation. (They do significantly
alter the
predicted GUT scale and coupling.)  

The Higgs multiplets form a
significant exception, however.  
The Higgs particle in the standard
model is {\it not\/} part of a complete $SU(5)$ multiplet at
low energies; its color triplet partner is capable of mediating
proton decay, and must be extremely heavy.  There
is no convincing theoretical
explanation of why it should be -- this is one aspect
of the gauge hierarchy problem.  In passing to the supersymmetric
version of the minimal unified model one must add the fermion
partners of this standard model doublet.  In fact, for slightly
subtle reasons, one actually must add
two such Higgs complexes, for it is
impossible to maintain supersymmetry with only a single Higgs field
giving masses to both up and down quarks.

\FIG\susyrun{Modification of the previous figure, if one includes the
effect of adding the minimal set of particles necessary to implement 
supersymmetry near the scale of electroweak symmetry breaking.
The constraint implied by unification is now satisfied.} 

Thus in the minimal
supersymmetric model one must add quite a few fields that do not
form complete $SU(5)$ multiplets and do affect the constraint among
low-energy couplings.  Remarkably, when this is done the
modified prediction agrees with the accurate modern experiments, as is
shown
in Figure \susyrun .

If we take this agreement at face value, as an indication for the
effect of virtual supersymmetry, it augurs a bright future for
experimental high energy physics.  If supersymmetry is to fulfill its
natural role in ameliorating the gauge hierarchy problem, it cannot be
too badly broken.  Specifically, if the the cancellation between
virtual particles and their supersymmetric partners is not to generate
corrections to the Higgs mass which are formally larger than that mass
itself, the generality of superpartners cannot be much heavier than
$M_W /\alpha~\approx 10~{\rm TeV}$.  Some are expected to be
considerably lighter.  Thus they fall within the range of foreseeable
accelerators.  Also, although supersymmetric unification raises the
GUT scale and thus decreases the rate for proton decay by exotic gauge
boson exchange, it does not do so by an enormous factor.  The
predicted range of rates, although safe from existing bounds, does not
seem hopelessly out of reach.

\section{Third Conceptual Revolution: The Early Universe Opens
to View}

The position of very early universe cosmology just prior
to the discovery of asymptotic freedom is well conveyed in
Weinberg's classic text [\wein ] (1972):

``However, if we look back a little further, into the first 0.0001 sec
of cosmic history when the temperature was above $10^{12}~$K, we
encounter theoretical problems of a difficulty beyond the range of
modern statistical mechanics.  At such temperatures, there will be
present in thermal equilibrium copious numbers of strongly interacting
particles -- mesons, baryons, and antibaryons -- with a mean
interparticle distance less than a typical Compton wavelength.  These
particles will be in a state of continual mutual interaction, and cannot
reasonably be expected to obey any simple equation of state.''

This pessimistic picture
changed overnight when the discovery of asymptotic freedom.
Instead of being mysterious and intractable, matter
at extreme temperatures and densities
becomes weakly interacting and its
equation of state simply calculable.  This
development, together with the ideas of unification just mentioned,
opened up a vast new field of investigation.
{\it It becomes possible to
make reasonable guesses for the behavior of matter under much more
extreme conditions\/}
than the mere $10^{12}~$K mentioned by Weinberg, and
to calculate the consequences of various unification scenarios for
cosmology with
some confidence.

\REF\kt{For a review of very early universe cosmology
see E. Kolb and M. Turner, {\it The Early Universe\/}
(Addison Wesley, Redwood City 1990).}

What has emerged from this opening [\kt ]?
There is now at least one plausible
scenario, based on baryon-number violating interactions
at the grand unified scale,
for how the asymmetry between matter and antimatter could
have developed from a symmetric starting condition.  Much
recent work has been devoted to the possibility of developing
such an asymmetry at the weak scale, though
the viability of this idea is presently unclear.
Various unification models can be constrained cosmologically
(for example if they create stable domain walls, contain too many
massless neutrinos, or contain
stable particles that would be produced in the
big bang in sufficient abundance that they could not have
escaped notice, ...).  Most exciting, plausible
candidates for the ``dark matter'' or ``missing mass'' have emerged.
These are particles that -- given their existence --
one can calculate would have been produced
in the big bang in sufficient abundance to redress the mismatch
between the density of ordinary matter observed directly or inferred
from nucleosynthesis and the amount necessary to account for the
gravitational dynamics of galaxies and clusters.  It is also necessary,
of course, that the postulated particles would have escaped observation
to date.  It is remarkable that two specific
kinds of particles: axions as
mentioned above, and LSPs (lightest supersymmetric particles) which
arise naturally in supersymmetric unification as discussed below,
seem to fit the
bill.  Heroic experiments are proposed to search for these particles,
whose (very different) properties are reasonably definitely predicted.
It would be difficult to overstate the importance of
a positive detection.
Finally, an impressive circle of ideas
around inflation has developed.

The ultimate value of
these specific, very speculative ideas
can't yet be reliably assessed.
I believe, however, we can already conclude
that the once seemingly impenetrable
veil of ignorance described by Weinberg,
which appeared to separate us
from sensible scientific
contemplation of the earliest moments
of the big bang, will
never again seem
so formidable.  Truly ``we live in the age of the trembling of
the veil.''

{\bf Acknowledgments}:  S. Treiman for helpful advice on the manuscript;
and M. Alford, K. Dienes, C. Kolda, J. March-Russell, G. Jungman, R. Ball and
S. Forte for help with the figures.

\endpage

\refout

\endpage

\figout

\endpage

\end

\np



$$\eqalignno{
^{n}O_{\mu_1}^{V} \cdots _{\mu_n} &= i\ ^{n-2} S\ {\rm Tr} \
F_{\mu_{1} \alpha} \nabla_{\mu_2} \cdots \nabla_{\mu_{n-1}} F^{\alpha}
_{\mu_n} \cr
&{\rm -trace ~ terms,}&(33) \cr
^{n}O_{\mu_1}^{F \pm, o} \cdots _{\mu_n} &= {1 \over 2}\ i\ ^{n-1} S\
\overline{\psi} \gamma_{\mu_1} \nabla_{\mu_2} \cdots \nabla_{\mu_n} (1 \pm
\gamma_{5})\ \psi \cr
&{\rm -trace ~ terms,}&(34) \cr
^{n}O_{\mu_1}^{F \pm, a} \cdots _{\mu_n} &= {1 \over 2}\ i\ ^{n-1} S\
\overline{\psi} \gamma_{\mu_1} \nabla_{\mu_2} \cdots \nabla_{\mu_n} (1 \pm
\gamma_{5}) {1 \over 2}\ \lambda^{a} \psi \cr
&{\rm -trace ~ terms,}&(35) \cr
}$$

\end